\documentclass[runningheads]{llncs}
\usepackage[T1]{fontenc}
\usepackage{graphicx,verbatim}
\usepackage{amsmath}
\usepackage{graphicx,verbatim}
\usepackage{color}
\usepackage{url} 

\usepackage{xspace}
\usepackage{xcolor}

\begin{document}
\title{Numerical Uncertainty in Linear Registration:\\ An Experimental Study}
\author{
\textsuperscript{*}Niusha Mirhakimi\inst{1}\orcidID{0009-0005-9558-1817} \and
Yohan Chatelain\inst{2}\orcidID{0000-0001-7023-250X} \and
†Jean-Baptiste Poline\inst{1}\orcidID{0000-0002-9794-749X} 
\and †Tristan Glatard\inst{2}\orcidID{0000-0003-2620-5883}
}

\authorrunning{N. Mirhakimi et al.}

\institute{
McConnell Brain Imaging Centre, The Neuro, Faculty of Medicine and Health Sciences, McGill University, Montreal, Quebec, Canada 
\and
Krembil Centre for Neuroinformatics, Centre for Addiction and Mental Health, Toronto, Ontario, Canada\\
\textsuperscript{*}Corresponding Author: \email{niusha.mirhakimi@mail.mcgill.ca}\\
†Co–Lead Senior Authors
}

\maketitle              

\begin{abstract}
While linear registration is a critical step in MRI preprocessing pipelines, its numerical uncertainty is understudied. Using Monte-Carlo Arithmetic (MCA) simulations, we assessed the most commonly used linear registration tools within major software packages—SPM, FSL, and ANTs—across multiple image similarity measures, two brain templates, and both healthy control (HC, n=50) and Parkinson's Disease (PD, n=50) cohorts. Our findings highlight how linear registration tools and similarity measures influence numerical stability. Among the evaluated tools and with default similarity measures, SPM exhibited the highest stability. FSL and ANTs showed greater and similar ranges of variability, with ANTs demonstrating particular sensitivity to numerical perturbations that occasionally led to registration failure. Furthermore, no significant differences were observed between healthy and PD cohorts, suggesting that numerical stability analyses obtained with healthy subjects may be generalizable to clinical populations. Finally, we also demonstrated how numerical uncertainty measures may support automated quality control (QC) of linear registration results. Overall, our experimental results characterize the numerical stability of linear registration experimentally and can serve as a basis for future uncertainty analyses.


\end{abstract}
\section{Introduction}
Neuroimaging preprocessing steps, including linear and non-linear registration, and segmentation, are sensitive to subtle numerical perturbations in data, pipeline configuration, or hardware environment~\cite{kiar2020comparing,pepe2023numerical,vila2024impact}. In some cases, these instabilities propagate to higher-level analyses, such as parcellation-based connectivity mapping~\cite{kiar2020comparing}, impacting derived findings.  

Linear registration is a critical step in the vast majority of neuroimaging preprocessing pipelines, commonly formulated as an optimization problem that aims to align a subject’s brain image to a common template. Small numerical errors introduced during optimization may steer the solution towards local minima, or prevent convergence. Therefore, understanding how numerical errors impact linear registration is crucial. The works in~\cite{salari2021accurate,vila2024impact} demonstrated that MCA can effectively simulate the effects of software updates and hardware-induced variability, including for FSL’s FLIRT registration tool. 

In this paper, we studied the numerical stability of widely-used linear registration tools. We also investigated the feasibility of using MCA-derived measures for automated QC. Our findings offer insights into tool selection and support the development of more reproducible and reliable preprocessing pipelines.

\section{Material and Methods}
\subsection{Monte-Carlo Arithmetic}
MCA is a commonly-used technique to investigate numerical instability in real-life software code bases~\cite{parker1997monte}. It utilizes randomness to simulate the effect of finite precision in floating-point (FP) operations, mimicking the effect of rounding errors and catastrophic cancellation~\cite{parker1997monte}. In this study, we used the random rounding perturbation mode, which injects controlled amounts of noise into the output of FP functions using the following perturbation:
\begin{equation}\label{mca_eq}
random\_rounding(x \circ  y)=round(inexact(x \circ y)),
\end{equation}
where $x$ and $y$ are FP numbers that represent the function’s inputs, $\circ$ is an arithmetic operation, and \emph{inexact} is a random perturbation at a given virtual precision:
\begin{equation}\label{inexact}
inexact(z)=z+2^{e_z-t}\epsilon,
\end{equation}
where $z$ is the original FP value, $e_z$ is the exponent of $z$’s FP representation, $t$ is the virtual precision, and $\epsilon$ is a random variable uniformly distributed in (-0.5, 0.5). 


The Verificarlo~\cite{denis2015verificarlo} and Verrou~\cite{fevotte2016verrou} frameworks implement random rounding to assess numerical stability. Verificarlo is an MCA tool built on the LLVM compiler infrastructure, supporting a wide range of languages such as C, C++, and Fortran. Verrou, in contrast, uses dynamic binary instrumentation via Valgrind, allowing perturbation of FP operations at runtime. Due to its lower computational overhead, we employed Verificarlo as the primary framework and used Verrou to validate a key result. Specifically, we utilized \textit{fuzzy-libm}~\cite{salari2021accurate}, a lightweight Verificarlo backend that perturbs only the outputs of standard mathematical functions from the \texttt{libm} library (e.g., \texttt{exp}, \texttt{sin}, \texttt{log}). While \textit{fuzzy-libm} is suitable for assessing the stability of programs relying heavily on \texttt{libm}, Verrou provides a more general solution by perturbing all FP operations (e.g., \texttt{+}, \texttt{-}, \texttt{$\times$}, \texttt{$\div$}) during execution.

\subsection{Numerical uncertainty metric}

The Framewise Displacement (FD) metric was introduced as a single measure to characterize head movement through a subject's time series~\cite{power2012spurious,power2014methods}. FD summarizes motion parameters into a displacement measured at a distance of 50 mm from the origin, approximating the mean radius of an adult brain:
\begin{equation}\label{fd_i}
    FD_i=\lVert t_i \rVert+50.\left(\frac{\pi}{180}\right)\lVert r_i \rVert,
\end{equation}
where $FD_i$ represents the framewise displacement for a transformation labeled $i$, $t_i$ is the translation vector in mm, $r_i$ is the rotation vector of Euler angles in degrees, and $\lVert.\rVert$ is the Euclidean norm. We assume that rotation and translation parameters correlate with shear and scaling parameters, which enables us to summarize affine 12-parameter registration with the FD measure.

To study how the FD varies under numerical perturbation for a given subject, we used the standard deviation (SD) of FD across the MCA runs as a measure of numerical uncertainty. Since the distributions of SD values were not normally distributed (see p-values from the Shapiro–Wilk test in Table~\ref{tab:pvalues}), we used non-parametric hypothesis testing to compare across tools and similarity measures.



\subsection{Similarity measures and optimization methods}

Three linear registration tools were evaluated in this study. FMRIB’s Linear Image Registration Tool (FLIRT), part of the FMRIB Software Library (FSL)~\cite{smith2004advances}, antsRegistrationSyN.sh script in Advanced Normalization Tools (ANTs)~\cite{avants2014insight}, and Statistical Parametric Mapping's (SPM) spm\_affreg function. Both FSL and ANTs employ multiresolution optimization strategies: registration begins with a coarse alignment at 8~mm resolution and is progressively refined through stages at 4~mm, 2~mm, and finally 1~mm~\cite{jenkinson2002improved,avants2011reproducible}. FSL uses the correlation ratio (CR) as its default similarity measure, whereas ANTs uses mutual information (MI). SPM adopts a fundamentally different optimization approach, using a Bayesian framework to estimate the affine transformation by iteratively incorporating prior knowledge and minimizing alignment errors~\cite{ashburner1997incorporating,ashburner2014spm12}. SPM employs the Sum of Square Differences (SSD) as its default similarity measure.

\subsection{Dataset and Preprocessing}
Fifty subjects with PD (age: $61.63 \pm 7.29$ years; 22 female; UPDRS3\_OFF: $21.55 \pm 11.42$) and fifty HC subjects (age: $62.17 \pm 10.48$ years; 24 female; UPDRS3\_OFF: $1.19 \pm 2.21$) were randomly selected from the baseline session of the  Parkinson's Progression Markers Initiative (PPMI) dataset~\cite{marek2011parkinson}, an ongoing, multicenter observational study to identify PD biomarkers. Hoehn and Yahr (HY) scores were available for 49 of the 50 PD subjects: 25 were classified as stage 1, 23 as stage 2, and 1 as stage 3. In the HC group, HY scores were available for 48 subjects: 47 were scored as stage 0, and 1 as stage 1.


T1-weighted images for both cohorts were processed using FSL's RobustFOV to remove the neck and BET to strip the skull. The resulting brain-extracted images were then used throughout the study as inputs to different linear registration pipelines. Each image was registered twice to a template for a given pipeline: first, using the standard (unperturbed) registration tools (referred to as the "IEEE registration"), and second, ten times using the MCA-perturbed registration tools (referred to as the "MCA registrations"). QC was performed manually on the registered images to ensure proper alignment: key brain structures, including the ventricles, corpus callosum, cerebellum, basal ganglia, brainstem, and Sylvian fissure, were assessed for alignment, along with the overall edge alignment of the brains to the template across the sagittal, coronal, and axial planes. All registered images, either IEEE or MCA, were visually inspected and labeled as "failed" or "passed." However, a subject's overall QC classification was determined separately for each registration pipeline, solely based on the quality of their corresponding IEEE registration. This approach enables exploring the potential of using MCA-derived features for automatic QC, specifically to distinguish between subjects who passed and failed QC based on the IEEE standard for each template and tool combination. Details regarding the templates and software versions used in this study are provided in Supplementary~\ref{supp:exp_setup}.

\section{Results}
\subsection{Numerical uncertainty can reach magnitudes comparable to template resolution or in-scanner head motion}
An examination of the median SD of FD across passed QC subjects for the three registration tools indicates generally low values: SPM ($1.7 \times 10^{-9}$mm), FSL (0.05 mm), and ANTs (0.07 mm). However, in ANTs and FSL, some subjects exhibited SD values greater than 0.2 mm—with a substantial portion of these observed for the MI and NMI similarity measures in FSL—and several cases exceeded 1 mm, surpassing the resolution of the registration template. In neuroimaging studies, mean displacements exceeding 0.2 mm within a subject are typically considered substantial motion artifacts~\cite{gu2015emergence}, as they represent a significant portion of the voxel dimension. These findings reveal that numerical instability in linear registration can, in some instances, introduce spatial variability comparable in magnitude to subject motion during a recording session. Table~\ref{tab:mean_sd} presents the mean SD per pipeline, highlighting that failed QC subjects consistently exhibit higher average variability in all pipelines.


\subsection{SPM is the most stable tool, FSL and ANTs are comparable}
\begin{figure}[htbp]

    \begin{minipage}[t]{0.5\textwidth}
        \includegraphics[width=\linewidth,height=5cm]{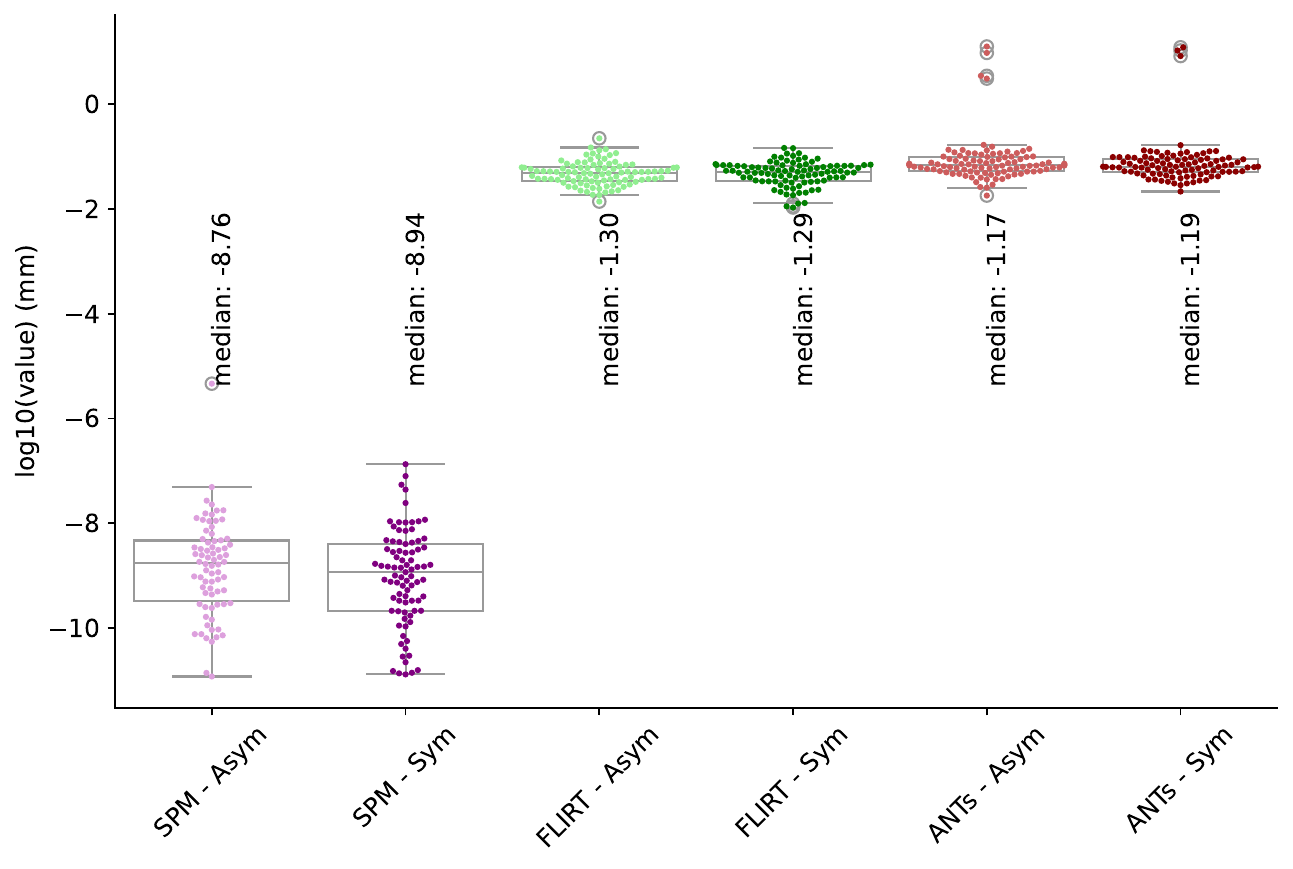}
        \small{(a) Comparison across linear registration tools (SPM, FSL, and ANTs) and templates (symmetric and asymmetric), using their default similarity measures }
        \label{fig:sd_fd_all}
    \end{minipage}
    \hfill
    \begin{minipage}[t]{0.47\textwidth}
        \raisebox{1.8em}{
            \includegraphics[width=\linewidth,height=4.cm]{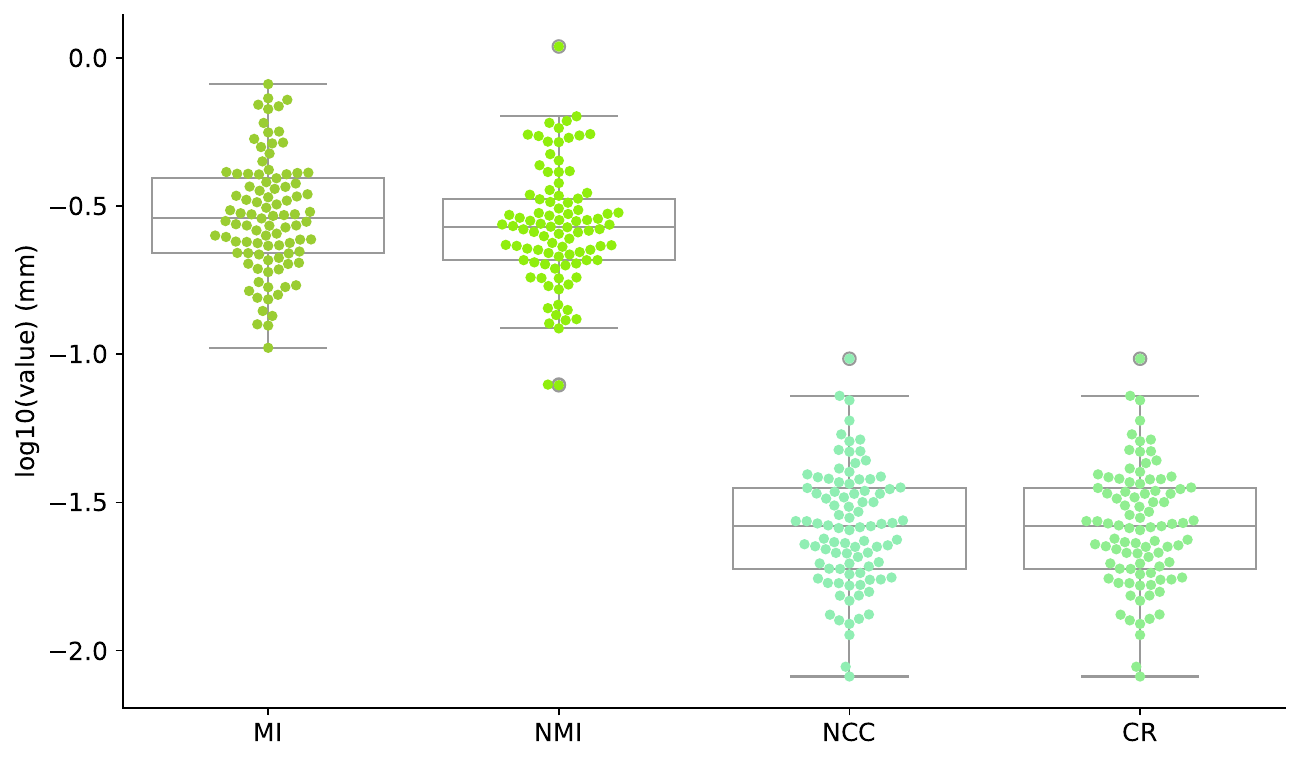}
        }
        \small{(b) Comparison of linear registration across similarity measures using FSL registered to the asymmetric template.} 
    \end{minipage}

    \caption{\textbf{Standard deviation of framewise displacement across MCA runs for each passed QC subject and registration pipeline.}}
    \label{fig:sd_fd_figure}
\end{figure}

We computed the SD of FD in MCA runs per subject for each registration tool with its default settings. Comparing SD distributions revealed notable differences in numerical uncertainty associated with each tool, confirmed by Friedman tests ($p \approx 3 \times 10^{-27}$ for registering to the asymmetric template and $p \approx 3 \times 10^{-29}$ for registering to the symmetric template). While FSL and ANTs exhibit comparable median variability, SPM demonstrates significantly greater numerical stability.  A closer examination of Figure~\ref{fig:sd_fd_figure}(a) illustrates that despite the similar range of uncertainty for ANTs and FSL, ANTs produces outliers—subjects with high variability across MCA runs. We performed a detailed visual QC on all images registered using instrumented tools and discovered a unique sensitivity in ANTs: 4 subjects who previously passed QC under unperturbed conditions failed under MCA-perturbed conditions. These failures directly contribute to the observed outliers.

\subsection{Similarity measure significantly affects numerical stability}
To isolate the impact of similarity measure selection, we conducted further experiments using FLIRT. Subjects were registered to the asymmetric template using various similarity measures: SSD, Normalized Cross Correlation(NCC), CR, MI, and Normalized Mutual Information (NMI)—as described in previous works~\cite{roche1998correlation,deserno2010fundamentals,hill2001medical}. SSD results were excluded since more than half of the unperturbed registrations failed. These failures were characterized by implausible transformation matrices, like a large scaling factor, that still produced low loss values. This is a known limitation of multi-resolution optimization schemes~\cite{jenkinson2001global} 
Analyzing the SD of FD distributions across MCA runs revealed that the choice of similarity measure significantly affects numerical stability (Friedman test: $p \approx 8 \times 10^{-54}$). As shown in Figure~\ref{fig:sd_fd_figure}(b), NCC and CR yielded more stable registrations than MI and NMI among subjects who passed QC. However, this result may not generalize to other registration tools, as discussed in the supplementary experiment (Supplementary~\ref{appendix:MI_comparison}).


\subsection{Numerical uncertainty metrics hold promise for automated QC}
\begin{figure}
    \centering

    \begin{minipage}[t]{0.3\textwidth}
        \centering
        \includegraphics[width=0.9\linewidth]{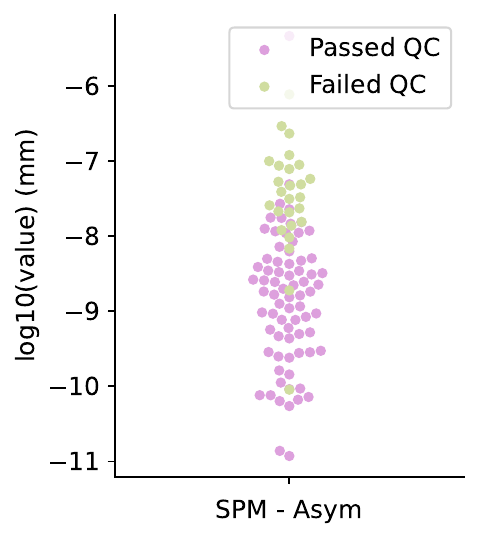}
        \small (a)
        \label{fig:sd_default}
    \end{minipage}
    \hfill
    \begin{minipage}[t]{0.3\textwidth}
        \centering
        \includegraphics[width=0.9\linewidth]{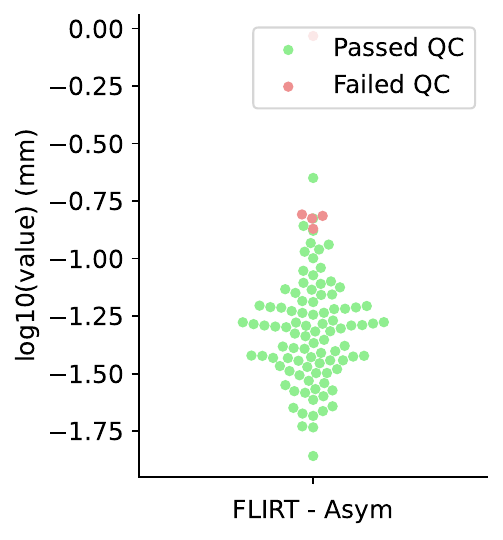}
        \small (b)
    \end{minipage}
    \hfill
    \begin{minipage}[t]{0.3\textwidth}
        \centering
        \includegraphics[width=0.9\linewidth]{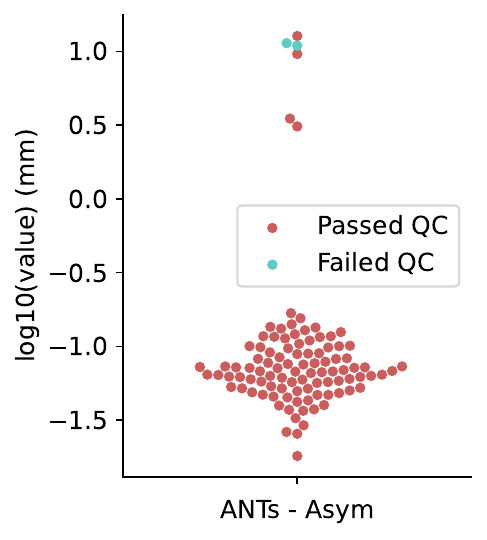}
        \small (c)
        \label{fig:fig:sd_costfunctions}
    \end{minipage}

    \caption{\textbf{Comparison of the standard deviation of framewise displacement between passed and failed QC subjects.} (a) SPM, (b) FLIRT, and (c) ANTs.}
    \label{fig:fd_qc}
\end{figure}

We investigated the potential of numerical uncertainty measures for automated QC in preprocessing pipelines. A consistent pattern emerges in the standard deviation of framewise displacement across pipelines (Figure~\ref{fig:fd_qc}), with failed cases exhibiting greater variability than those that passed QC. The distinction was most pronounced in SPM, where the number of failed and passed subjects allowed for observation of two separate distributions. Due to the imbalance in sample sizes, we employed one-class classification models trained on passed QC subjects and treated failed cases as anomalies. Since all test cases were known failures, we report recall as the evaluation metric available in Table~\ref{tab:recall}. While this setup is not intended as a definitive classification approach, it serves as a proof of concept, illustrating that numerical uncertainty measures capture meaningful aspects of registration quality and may serve as promising features in future automated QC systems.


\begin{table}
\caption{\textbf{Evaluation of novelty detection methods for identifying failed subjects.} Recall scores of three novelty detection methods in distinguishing passed vs. failed QC subjects based on the SD of FD. No threshold optimization or training was applied. Each linear registration pipeline processed the same 100 images; the number of passed cases is \(100 - N\), where \(N\) is the number of QC failures.}

\label{tab:recall}
\begin{center}
\begin{tabular}{|p{1.7cm}p{1.7cm}p{1.7cm}p{0.75cm}p{1.7cm}p{1.7cm} p{1.5cm}|}
\hline
   \textbf{Template} &   \textbf{Software} & \textbf{Similarity Measure} &    \textbf{N} & \textbf{95th Quantile} &   \textbf{KDE (5\%)} &  \textbf{1-Class SVM} \\
\hline

                                    Asym &  FSL &  NCC & 6 &  0.83 &   0.83 & 1.0 \\ 
                                    
                                      Asym & FSL  &  NMI &  5 & 0.8 &  1.0
                                        & 0.4 \\ 
                                      Asym&  FSL &  MI &  7 &  1.0 &  1.0 & 1.0\\ 
                                      Asym&  FSL &  CR &  5 & 1.0 &  1.0 & 0.2\\ 
                                      Asym&  ANTs &  MI &   2 &  1.0 & 1.0 & 1.0\\ 
                                      Asym&  SPM &  SSD &  26 &   0.73 &  0.73 & 1.0\\ 
                                      sym&  FSL &  CR &  6 &  0.5 &  0.5 & 0.5\\ 
                                    Sym &   ANTs &  MI &  2 &  1.0 &               1.0 & 1.0\\ 
                                      Sym&  SPM &  SSD &  15 &  0.6 &  0.6 &  1.0 \\
\hline
\end{tabular}
\end{center}
\end{table}

\subsection{Cohort and template choices show no statistically significant impact on numerical stability}
We investigated whether cohort (PD vs. HC) or template choice (asymmetric vs. symmetric) significantly affects the numerical stability of linear registration, as measured by the SD of FD across MCA runs. For cohort effects, Mann–Whitney U tests revealed no statistically significant differences in numerical uncertainty between PD and HC subjects across all registration pipelines (Table~\ref{tab:pvalues}). Regarding template choice, Wilcoxon signed-rank tests comparing registrations to symmetric and asymmetric templates showed no significant differences for FSL ($p \approx 0.479$) and ANTs ($p \approx 0.329$). While SPM showed a marginally significant difference ($p \approx 0.0194$), this was likely driven by a single outlier, with minimal visual difference observed in the overall distributions. These findings, together, suggest that neither cohort differences nor template choice substantially influence numerical stability in the evaluated registration tools.

\section{Discussion and Conclusion}

This work investigates the numerical uncertainty of linear registration. While focusing on a single dataset cannot fully disentangle the effects of software choice, similarity measures, demographics, and template selection, it nonetheless serves as an initial case study, highlighting the importance of investigating numerical uncertainty as an overlooked source of variability. Numerical stability is crucial for a pipeline’s robustness, and this study sets the stage for more comprehensive evaluations and future efforts to develop more reliable neuroimaging workflows.

Numerical variability in linear registration can be significant, in some cases comparable to template resolution and head movement in the scanner, suggesting it potentially leads to substantial discrepancies in preprocessing and downstream analyses for some subjects.

The numerical stability of linear registration is strongly software-dependent. Among the tools evaluated, SPM exhibited markedly greater stability, while FSL and ANTs showed heightened sensitivity to numerical perturbations. We considered two potential explanations for this difference: either SPM was not properly instrumented with Verificarlo, or its underlying optimization strategy contributes to its stability. To rule out instrumentation issues, we conducted supplementary validation experiments (Supplementary~\ref{supp:spm_validate}). We made sure Octave was instrumented as expected, and additionally, we instrumented SPM with the Verrou framework. The comparable variability patterns observed across both tools suggest that the instrumentation was effective, supporting the hypothesis that SPM’s increased numerical stability stems from its fundamentally different optimization approach.

Multi-resolution approaches are predicated on the assumption that the minima identified at lower resolutions are sufficiently close to the global minima at higher resolutions, which is necessary for convergence~\cite{hill2001medical}. However, there is no guarantee that this assumption holds, as local minima can shift across different resolutions. Additionally, the processes of subsampling and interpolation introduce noise, in which multi-resolution approaches not only fail to simplify the optimization landscape but also introduce new local optima, adding complexity to the problem~\cite{jenkinson2001global}. Multi-resolution methods involve complex subsampling and interpolation steps that can amplify small perturbations, potentially leading to suboptimal solutions.

SPM employs a Bayesian optimization framework that incorporates prior knowledge about variability in head shape and size, using a Maximum A Posteriori approach to enhance robustness and convergence speed~\cite{ashburner1997incorporating}. This method is particularly advantageous when dealing with low-quality data, as it reframes the optimization objective to not only maximize image similarity but also penalize deviations from expected parameter values based on prior distributions. This strong regularization may explain SPM’s observed numerical stability and its resilience to small perturbations.

The choice of similarity measure influences numerical stability, and this effect varies across software. FSL was selected to evaluate similarity measures, as it supports both SSD (used by SPM) and MI (used by ANTs). ANTs does not support CR, and SPM’s spm\_affreg hardcodes SSD. A comparison of MI in FSL and ANTs (Supplementary~\ref{appendix:MI_comparison}) showed that stability depends on both the cost function and the tool. This suggests that each tool should be evaluated with multiple similarity measures, as optimization strategies may interact with cost functions. Further investigation is needed to understand ANTs’ instability and assess broader configurations.

 Given that linear registration tends to struggle in the presence of brain atrophy~\cite{dadar2018comparison}, that patient data often suffers from motion-related artifacts\cite{gilmore2021variations}, and that poor image quality can amplify numerical instability during processing~\cite{chatelain2022pytracer,salari2021accurate}, the PD cohort was expected to exhibit reduced numerical stability. However, no significant differences were observed between the PD and control groups, suggesting that numerical uncertainty findings from healthy populations may be generalizable to pathological cohorts. Nevertheless, we hypothesize that even small perturbations introduced during registration may propagate through the full preprocessing pipeline, potentially influencing downstream metrics such as cortical thickness and altering observed effect sizes between groups.

 This study demonstrated that numerical variability measures hold promise for integration into automated QC algorithms within preprocessing pipelines, potentially enhancing the reliability of neuroimaging workflows. We used variability in FD as a proxy for numerical uncertainty and showed that subjects who failed QC exhibited higher variability. As a future direction, an alternative metric could be developed based on the Anatomical Fiducials Registration Error, a method introduced by~\cite{lau2019framework}, which uses 32 anatomical fiducial points identified on brain scans to assess registration accuracy. Studying the variability of these fiducial points across perturbed runs may offer a more localized and anatomically meaningful estimation of numerical uncertainty.

\bibliographystyle{splncs04}
\bibliography{main}
%





\clearpage
\appendix
\section{Supplementary Material}
\addcontentsline{toc}{section}{Supplementary Material}
\renewcommand{\thefigure}{A\arabic{figure}}
\renewcommand{\thetable}{A\arabic{table}}

\subsection{Experimental Setup} \label{supp:exp_setup}
In this study, the symmetric (Sym) and asymmetric (Asym) versions of the MNI152NLin2009c template, each with a resolution of 1 mm were exclusively used. The templates are accessible through the TemplateFlow website at~\url{https://www.templateflow.org}.
The Dockerized formats of SPM12, ANTs v2.5.0, and FSL v6.0.4 were utilized throughout the study. Docker recipes for both unperturbed and perturbed software versions are available in this GitHub repository: ~\url{/www.github.com/mirhnius/mca_linear_registration}.

\subsection{Verifying SPM perturbation process: Octave and Verrou instrumentation} \label{supp:spm_validate}
To verify that Octave utilizes the \texttt{libm} (math) library and ensure compatibility with Verificarlo, we selected common mathematical functions—such as \texttt{sin}, \texttt{cos}, \texttt{exp}, and \texttt{log}—and evaluated their outputs for a fixed set of inputs. We compared the results between the standard and instrumented versions of Octave. The perturbed outputs showed consistent variation across 10 runs, confirming that Octave was successfully instrumented with Verificarlo and relies on \texttt{libm}.

Although this experiment validated Verificarlo’s integration with Octave, uncertainty remained regarding its proper instrumentation of SPM. To address this, we used Verrou, a more general framework that applies runtime perturbations beyond \texttt{libm}. Verrou was configured in random rounding mode and used to instrument SPM. We performed 10 registration runs per subject to the asymmetric template. Comparing the SD of FD between Verrou and Verificarlo runs revealed a similar range of variability (Figure~\ref{fig:supplementary_comp}(a)). This consistency suggests that the high numerical stability observed in SPM, relative to FSL and ANTs, likely stems from its robust optimization strategy.

\subsection{Assessing numerical stability of mutual information-based linear registration in FSL and ANTs}\label{appendix:MI_comparison}
\begin{figure}[htbp]
    \centering

    \begin{minipage}[t]{0.45\textwidth}
    \begin{center}
        \includegraphics[width=0.55\linewidth,height=3.2cm]{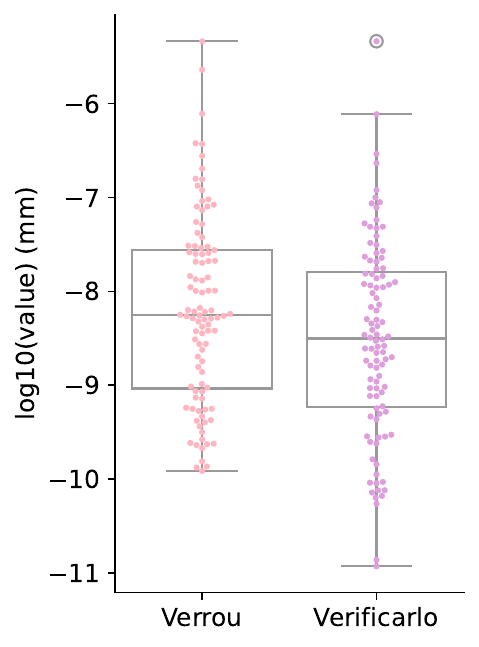}
    \end{center}
        
        
        \small{(a) Numerical variability in SPM's linear registration under two perturbation frameworks.}
        
    \end{minipage}
    \hfill
    \begin{minipage}[t]{0.45\textwidth}
    \begin{center}
    \includegraphics[width=0.55\linewidth,height=3.2cm]{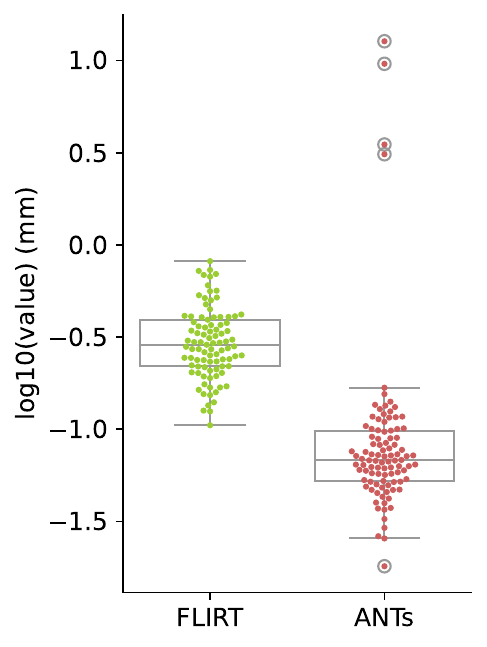}
    \end{center}
        \small{(b) Numerical variability with MI similarity measure in FLIRT and ANTs linear registration.} 
    \end{minipage}

    \caption{\textbf{Supplementary comparisons of the standard deviation of framewise displacement of passed QC subjects across MCA runs.}}
    \label{fig:supplementary_comp}
\end{figure}

 A comparative analysis of ANTs and FSL, both using MI as the similarity measure, indicates that ANTs generally exhibits greater numerical stability (Wilcoxon signed-rank test, $p \approx 3 \times 10^{-12}$). This is supported by lower variability in the SD of FD as shown in Figure~\ref{fig:supplementary_comp}(b). This comparison underscore that numerical stability is influenced not only by the choice of similarity measure but also by the specific implementation within each software tool. Despite the general stability of ANTs with the MI cost function, these failures underscore the need for ongoing investigations into the numerical stability of ANTs, and it remains essential to determine the origin of this sensitivity.

\subsection{Tables and Plots}

\begin{table}[htbp]
    \centering
    \caption{Comparison of Mean SD of Framewise Displacement for Passed and Failed QC Subjects across Templates and Software.}
    \label{tab:mean_sd}
    \renewcommand{\arraystretch}{1.2}
    \begin{tabular}{|p{1.6cm}p{1.7cm}p{2cm}p{2.3cm}p{2cm}|}
        \hline
        \textbf{Template} & \textbf{Software} & \textbf{Similarity\newline Measure} & \textbf{Mean SD\newline Passed} (mm)& \textbf{Mean SD\newline Failed} (mm)\\
        \hline
                                    Asym&  FSL &  NCC &$2.9 \times 10^{-2}$&$1.8\times 10^{1}$\\
                                      Asym&  FSL &  NMI &$3.1 \times 10^{-2}$&$2.4\times 10^{1}$\\
                                      Asym&  FSL &  MI &$3.2 \times 10^{-1}$&$3.2\times 10^{1}$\\
    
                                      Asym&  FSL &  CR &$5.5 \times 10^{-2}$&$3.1\times 10^{-1}$\\
                                      Asym&  ANTs &  MI  &$3.6 \times 10^{-1}$&$1.1\times 10^{1}$\\
                                      Asym&  SPM &  SSD  &$6.7 \times 10^{-8}$&$8.6 \times 10^{-8}$\\
                                      
                                      Sym &  FSL &  CR &$5.5 \times 10^{-2}$&$3.8\times 10^{-1}$\\
                                    Sym &  ANTs &  MI  &$3.9 \times 10^{-1}$&$1.1\times 10^{1}$\\                                  
                                      Sym &  SPM &  SSD &$6.1 \times 10^{-9}$&$9.7 \times 10^{-8}$\\

\hline
\end{tabular}
\end{table}

\begin{table}[htbp]

   \caption{\textbf{Normality and group comparison Tests on the Standard deviation of framewise displacement.} Both Shapiro–Wilk and Mann–Whitney U tests were applied on passed QC subjects to assess normality of distributions and compare PD vs. HC groups, respectively.}
   
    \label{tab:pvalues}
\begin{center}
\begin{tabular}{|p{1.7cm}p{1.7cm}p{1.7cm}p{2.2cm}p{2cm}|}
\hline

           \textbf{Template} & \textbf{Software} & \textbf{Similarity\newline Measure} & \textbf{Normality}\newline (p-value) & \textbf{PD vs HC}\newline (p-value) \\
\hline

                                    Asym&   FSL &  NCC &  $1.6 \times 10^{-13}$ &  0.280 \\ 
                                     Asym &   FSL &  NMI &    $2.6 \times 10^{-7}$ &  0.092 \\ 
                                      Asym&  FSL &  MI &   $3.1 \times 10^{-14}$  &  0.940 \\ 
     
                                      Asym&   FSL &  CR &   $3.9 \times 10^{-7}$  &  0.128\\ 
                                      Asym&  ANTs &  MI &   $8.3 \times 10^{-21}$  &  0.165\\
                                      Asym&  SPM &  SSD &   $4.3 \times 10^{-19}$  &  0.738 \\ 

                                      Sym&  FSL &  CR &     $3.2\times 10^{-4}$&  0.829 \\ 
                                    Sym &  ANTs &  MI &   $6.4 \times 10^{-21}$ & 0.400\\ 
                                      Sym&  SPM &  SSD &  $1.0 \times 10^{-17}$  &  0.442 \\ 

\hline
 
\end{tabular}
\end{center}
\end{table}

\end{document}